\documentclass[article,shortnames,nojss]{pseudo_jss}

\usepackage{algorithm}
\usepackage{algpseudocode}
\usepackage{subfig}
\usepackage{booktabs}
\usepackage{setspace}
\def\+{\discretionary{}{}{}} 

\author{Ulrich Matter \\University of Basel}
\title{\pkg{RWebData}: A High-Level Interface to the Programmable Web}

\Plainauthor{Ulrich Matter} 
\Plaintitle{RWebData: A High-Level Interface to the Programmable Web} 

\Abstract{The rise of the programmable web offers new opportunities for the empirically driven social sciences. The access, compilation and preparation of data from the programmable web for statistical analysis can, however, involve substantial up-front costs for the practical researcher. The \proglang{R}-package \pkg{RWebData} provides a high-level framework that allows data to be easily collected from the programmable web in a format that can directly be used for statistical analysis in \proglang{R} \citep{R_2013} without bothering about the data's initial format and nesting structure. It was developed specifically for users who have no experience with web technologies and merely use \proglang{R} as a statistical software. The core idea and methodological contribution of the package are the disentangling of parsing web data and mapping them with a generic algorithm (independent of the initial data structure) to a flat table-like representation. This paper provides an overview of the high-level functions for \proglang{R}-users, explains the basic architecture of the package, and illustrates the implemented data mapping algorithm.
}

\Plainkeywords{R, programmable web, big public data, web api, rest} 

\Address{
  Ulrich Matter\\
  Faculty of Business and Economics\\
  University of Basel\\
  Peter Merian-Weg 6\\
  4002 Basel, Switzerland\\
  E-mail: \email{ulrich.matter@unibas.ch}\\
  Website: \url{http://wwz.unibas.ch/matter}
}

\usepackage{hyperref}
\usepackage{breakurl}
\usepackage{float}

\begin{document}


\section[Introduction]{Introduction}
\label{sec:rwebdata_intro}

Digital data from the Internet has in many ways become part of our daily lives. Broadband facilities, 
the separation of data and design, as well as a broader adaption of certain web technology standards increasingly facilitate the integration of data  across different software applications and hardware devices over the web. The Internet is increasingly becoming a programmable web\footnote{I use the term programmable web synonymously for ``Semantic Web'' or ``Web of Data'' and as conceptually motivated by \citet{swartz_2013}.}, where data is published not only in HTML-based websites for human readers, but also in standardized machine-readable formats to be ``shared and reused across application, enterprise, and community boundaries'' \citep{w3c_2013}. The technical ecology of this 
programmable web consists essentially of web servers providing data in formats such as Extensible Markup Language (XML) via Application Programming Interfaces (APIs)\footnote{Web APIs are a collection of predefined HTTP requests and response messages to facilitate the programmatic exchange of data between a web server and clients.} to other server- or client-side applications (see, e.g., \citealt{nolan_lang2014} for a detailed introduction to web technologies for \proglang{R}-programmers). In the conceptual framework of this paper, APIs serve a dual function: they are the central nodes between different web applications and, at the same time, the central access points for researchers when they want to systematically collect and analyze data from the programmable web. More and more of these access points are becoming available every day. Moreover, this trend is likely to continue with the increasing number of people who have devices to access the Internet and the increasing amount of data recorded by embedded systems (i.e., sensors and applications in devices such as portable music players or cars that automatically feed data to web services).\footnote{The number of publicly accessible web APIs has grown from around 1,000 at the end of 2008 to over 10,000 at the end of 2013 \citet{progweb_2014}. \cite{idc_2014} estimate that the share of available digital data stemming from embedded systems will rise to 10\% by 2020 and will include up to 32 billion devices connected to the Internet. See also \cite{helbing_pournaras2015} for a discussion of how this `Internet of things' could be fostered on a crowd-sourced basis and employed for big-data analytics.}

The rise of the programmable web offers various new opportunities for data-driven
sciences. First, for the social sciences, it provides big data covering 
every-day human activities and interrelationships as people leave digital traces by using mobile devices and applications based on APIs. The systematic analysis of such data is likely to offer new insights in fields as diverse as economics, political science, and sociology (see \citealt{matter_stutzer2015plos} for a review of the arguments in the case of political economics and political science, as well as, e.g., \citealp{dodds2011, digrazie_etal2013, preis_2013flickr, barbera_2015, matter_stutzer2015} for primary research based on data collected via web APIs). In fact, web APIs might become an important domain of data sources for the evolving field of computational social science (see, e.g., \citealp{Lazer2009, cioffi2010, conte_etal2012, giles2012}). Second, researchers from all fields of science can provide their own prepared data via APIs as a resource to facilitate research, and the consequent aggregation of data sets from diverse disciplines offers new opportunities to create new `scientific data mash-ups' (\citealt{bell_2009}) and drive innovation. The provision and hosting of APIs by research institutes and scientists for other scientists is already common practice in the life sciences (see, e.g., the API to the NCBI Gene Expression Omnibus data repository, \citealt{ncbi_2014}). Third, the use of standardized protocols and data formats to exchange data over the web via APIs offers, in general, new approaches to facilitate reproducibility and replicability of research.

While the advantages of accessing the programmable web and integrating it into research projects are very promising, the practical utilization of this technology comes at a cost and demands that researchers possess a specific skill set and a certain knowledge of web technologies. The new \proglang{R}-package \pkg{RWebData} substantially reduces these costs by providing several high-level functions that facilitate the exploration and systematic collection of data from APIs based on a Representational State Transfer (REST) architecture\footnote{See \cite{richardson_2013} for an introduction to REST APIs.}. Moreover, the package contains a unified framework to summarize, visualize, and convert nested/tree-structured web data that works independently of the data's initial format (XML/RSS, JSON, YAML) as well as a framework to easily create \proglang{R} packages as client libraries for any REST API. The package is aimed at empirical researchers using \proglang{R} as their daily data analysis- and statistics tool, but who do not have a background in computer science or data science. In addition, several lower level functions of \pkg{RWebData} might also be useful for more advanced \proglang{R}-programmers who whish to develop client applications to interact with REST APIs. The package thus bridges the gap between the fundamental and very valuable \proglang{R}-packages that integrate several web technologies into the \proglang{R}-environment (see, e.g., \citealt{xml_2013, rcurl_2013,  rjson_2013, jsonlite_2014}) and the statistical analysis of the programmable web universe. 

The development and use of \pkg{RWebData} is motivated in Section~\ref{sec:rwebdata_motivation}, which explains the challenges of extracting data from the programmable web for statistical analysis and introduces the packages that enable \proglang{R}-users to work with web technologies. Section~\ref{sec:rwebdata_datamapping} presents the central data mapping strategy and algorithm developed in \pkg{RWebData} as well as an overview of the package's basic architecture. Section~\ref{sec:rwebdata_basics} illustrates the high-level functionality of \pkg{RWebData} for a typical user. In Section~\ref{sec:rwebdata_advanced}, more advanced examples show how \pkg{RWebData} can be used to write Open Source Interfaces/API client libraries. The concluding discussion in Section~\ref{sec:rwebdata_discussion} reviews the potential of \pkg{RWebData} for the empirically driven social sciences in terms of big public data access as well as the reproducibility and replicability of research based on data from web APIs.

\pkg{RWebData} is free, open source (under GLP-2 and GLP-3 license), and published on Bitbucket (\url{https://bitbucket.org/ulrich-matter/rwebdata}). It can directly be installed from the \proglang{R}-console (using the \pkg{devtools}-package; \citealt{devtools}) with the command \code{install_\+bitbucket('ulrich-matter/RWebData')}. This paper describes version 0.1 of \pkg{RWebData}, which requires \proglang{R} version 3.0.2 or a subsequent version.

\section[Motivation]{Motivation}
\label{sec:rwebdata_motivation}

How to extract data from the programmable web in a format suitable for statistical analysis? Particularly social scientists are often confronted with APIs whose initial purpose \emph{is not} to provide data for scientific research in order to make the researchers' lives easier, but to facilitate the integration of the data in dynamic websites and smartphone applications.\footnote{The vast majority of APIs are explicitly made for web developers and are hosted by companies or NGOs that have nothing to do with academic research (according to \citealt{santos_2012}, only about 195 of over 7,000 APIs listed in the ProgrammableWeb.com API-directory were related to `science' in 2012.).} In such a context, API server and API client share the problem domain, but do not share the same goal. This makes it per se harder to write an API client in order to compile data from the web API (see \citealt{richardson_2013} for a discussion of this issue).  The API query methods might, for example, not provide the length and breadth of data that the researcher is looking for using a single query, but only with a combination of several query methods. Importantly, the returned data are -- due to the initial purpose of most APIs -- usually not in a format that can be directly integrated in a statistical analysis. Researchers trying to compile data from such an API are thus confronted with two main challenges:
\begin{enumerate}
  \item The need to query and combine data from different API methods which is likely to involve many requests to the API and demands a basic knowledge of how to interact with an API from within the \proglang{R}-environment.
  \item The extraction and conversion of the data records from a nested web data format such as XML, JSON, or YAML to a table-like representation that can be directly used for statistical analyses.
\end{enumerate}

An additional challenge is to overcome the two issues above in a way that supports the reproducibility and replicability of research based on API data (including the pre-processing of raw data). As pointed out by \cite{matter_stutzer2015plos}, this is an increasingly important issue in the context of social science research which needs to be resolved, as the new digital data sources hinder the replicability of research owing to the present `unique purpose' of the new data sets.

\subsection{Interacting with REST web APIs}
In order to visit a certain website, we normally type a website's address in a web browser's address bar. The browser (a type of web \emph{client}) then sends a message to the web server behind that address, requesting a copy of the website and then parses and renders the returned website (usually a HTML document) in the browser window.
In REST terminology, the website's address is an \emph{URL} and the website which the URL locates is a \emph{resource}. The document that the server sends in response to the client's request is a \emph{representation} of that resource. Importantly, every URL points only to one resource and every resource should have only one URL (this is one of the REST principles and is referred to as \emph{addressability}). The message that the browser sends to the web server is a HTTP \emph{GET} request, a HTTP method that essentially requests a representation of a given resource.

REST web APIs work basically the same way as the website outlined above. The crucial difference is, that their design is intended for programmable clients (rather than humans using a web browser) and that they therefore consist of URLs pointing to resources that are not optimized for graphical rendering, but which instead contain the raw data (often in a format such as XML or JSON that facilitates the integration of the data in a HTML document). 

Interacting with a REST API from within the \proglang{R}-environment, therefore, means,  sending HTTP requests to a web server and handling the server's response with \proglang{R}. In addition to the basic functionality necessary for this, which are delivered in the \pkg{base}-package \citep{R_2013}, more detailed functions for HTTP requests are provided in packages such as \pkg{RCurl} \citep{rcurl_2013} and \pkg{httr} \citep{wickham2014_httr}. \pkg{RWebData} builds on the \pkg{libcurl}-based \pkg{RCurl} package which is designed to interact with web servers hosting a REST API. The implementation is focused on a robust download of resources that checks the received HTTP response for potential problems (e.g., unexpected binary content of the HTTP responses' body-entity) to make sure that the user does not have to specify anything other than the URL.

\subsection{Web data conversion}

Most functions related to the fitting and testing of statistical models as well as to exploratory data analysis and data visualization in the \proglang{R} computing environment work on data represented in data-frames (a table-like flat representation in which each row represents a data-record/observation and each column a variable describing these records).\footnote{Technically, a data-frame can contain other r-objects than just scalars (i.e., another data-frame). However, in the vast majority of applications in the context of statistical analysis and data visualization, data-frames are used for a table-like flat data representation.} Data-frames are probably the most common \proglang{R}-objects used by researchers when conducting an empirical analysis with \proglang{R}. Many input/output-functionalities in \proglang{R} serve to import (export) data into (from) \proglang{R} as data-frames.\footnote{See, i.e., data from CSV- and similar text files: \code{read.table()} in \cite{R_2013} or similar functions in \cite{readr_2015}, Microsoft Excel files: \code{read.xls()} in \cite{gdata_2014}, data from other statistical computing environments such as Stata: \code{read.dta()} in \cite{foreign_2014}, or data from ODBC-data bases \code{sqlQuery()} in \cite{rodbc_2013}.} However, when it comes to reading data from web APIs into \proglang{R}, this is not necessarily the case. The reason lies in the nature of web data formats which allow for more flexibility than solely a table-like data representation and come (in the context of web APIs) typically in a nested structure. While a conversion from one table in, e.g., a XML-document to a data-frame is straightforward (see, i.e., \code{xmlToDataFrame()} in \citealt{xml_2013}; see also \citealt{xml2r}), a conversion of a more complex XML-document with a nested data structure to a flat table-like data representation in \proglang{R} or any other computing environment is ex ante less clear and depends on the nature of the data and the purpose the data is converted for.\footnote{See, e.g., the classical problem of mapping a set of XML-documents to a relational data base scheme (RDBS; i.e., a set of linked tables) with or without knowing the scheme behind the XML-documents (\citealt{moh_2000re,  men_2001}). See also, in the \proglang{R} context, the different implementations of mapping JSON to \proglang{R}-objects in \cite{rjson_2013}, \cite{rjsonio_2013}, or \cite{jsonlite_2014} as well as a detailed discussion of this matter in \cite{ooms_2014}.} This is particularly the case for data provided by a non-scientific API that is explicitly made for web developers in order to integrate the data in dynamic websites or mobile applications and not for researchers to integrate the data in their analysis. Not surprisingly, there are different solutions offered in the \proglang{R}-environment to map web data formats such as JSON and XML to \proglang{R}-objects.\footnote{See, e.g., the CRAN Task View on web technologies and services (\url{http://cran.r-project.org/web/views/WebTechnologies.html}), the CRAN Task View on open data (\url{https://github.com/ropensci/opendata}), as well as the rOpenSci (\url{https://ropensci.org/packages/}) for an overview of popular \proglang{R}-packages that parse data from the web and map it to \proglang{R}-objects. Some contributions in this area also focus on the automated information extraction from traditional websites made for human interaction. Such methods are commonly summarized under the terms ``web scraping" or ``screen scraping".  See, e.g., the \proglang{R}-package \pkg{rvest} (\citealt{rvest}) for functions that facilitate web scraping as well as \cite{munzert_etal2015} for a practical introduction to web scraping with \proglang{R}.} Most packages represent web data as nested lists containing objects of different classes such as atomic vectors or data-frames. After converting the original web data to \proglang{R}-objects, the user, therefore, often has to extract and rearrange the data records of interest in order to obtain a data representation for statistical analysis. This process becomes even more complex, as queries to the same API method might provide slightly different results depending on the amount of detail embedded in the data records. In these cases, data extraction can become particularly costly if the data set that needs to be compiled depends on many API requests involving various API methods.

\section[Data mapping strategy and basic architecture]{Data mapping strategy and basic architecture}
\label{sec:rwebdata_datamapping}

Web data provided via REST APIs are typically in a format such as XML, JSON, or YAML and are not structured in tables containing data values in rows and columns, but are rather organized in a nested (tree-like) structure. However, independent of the data format and data structure, documents provided by a particular REST API have basic aspects in common. Based on these commonalities, this section presents a conceptual and terminological framework as well as a description of how this framework is used to develop the data mapping strategy applied in \pkg{RWebData} and serves as the foundation of the data mapping algorithm which will be outlined later on.
 
\subsection{Original data structure and data semantics}
From the researcher's point of view, the data sets provided by web APIs obey, independent of the raw data format and nesting structure, some very basic data semantics. For these data semantics, I use mainly the same terminology that is nicely outlined by \citet[p. 3]{wickham2014}: 

\begin{enumerate}
  \item ``A dataset is a collection of \emph{values}, usually either numbers (if quantitative) or strings (if qualitative)."
  \item ``Every value belongs to a \emph{variable} and an \emph{observation}."
  \item ``A variable contains all values that measure the same underlying attribute (like height, temperature, duration) across units."
  \item ``An observation contains all values measured on the same unit (like a person, or a day, or a race) across attributes."
\end{enumerate}

In addition, each observation and each variable belongs to a specific observation \emph{type}. Thus, in a multi-dimensional data set describing, for example, a city, observation types could be buildings or citizens. The following fictional XML example further illustrates this point.

\begin{Schunk}
\begin{Soutput}
<?xml version="1.0" encoding="UTF-8"?>
<firm>
  <employees>
    <employee>
      <firstName>John</firstName>
      <secondName>Smith</secondName>
    </employee>
    <employee>
      <firstName>Peter</firstName>
      <secondName>Pan</secondName>
    </employee>
  </employees>
  <shareholders>
    <shareholder>
      <ID>S1</ID>
      <Name>Karl Marx</Name>
    </shareholder>
    <shareholder>
      <ID>S2</ID>
      <Name>Bill Gates</Name>
    </shareholder>
  </shareholders>
  <firmName>MicroCapital Ltd</firmName>
  <firmID>123</firmID>
</firm>
\end{Soutput}
\end{Schunk}

The XML document contains a data set describing a firm. The variables ``firstName" and ``secondName" describe observations of type ``employee", while observations of type ``shareholder" are described by the variables ``ID" and ``Name". Finally, the variables ``firmName" and ``firmID" describe the firm itself, which builds another type of observation.
The following subsection illustrates how the data would be mapped according to the procedure implemented in \pkg{RWebData}.

\subsection{Mapping nested web data to data-frames}
The core idea behind the data mapping procedure in \pkg{RWebData} is built around the basic semantics outlined above. According to this system, documents returned from APIs can contain data describing observational units of one type or several different types. \pkg{RWebData} returns one data-frame for each observation type. Observations and variables belonging to different types are, thus, collected in different data-frames, each describing one type of observational unit. Some observations  might describe the document (or main type of observational unit) itself (i.e., metadata). These are collected in a separate data-frame. Tables~\ref{tab:mapping} (a-c) present the XML-document from the example above after mapping to data-frames according to the outlined mapping procedure. The next subsection explains how this process is implemented in \pkg{RWebData}.

\begin{table}[h]%
  \centering
  \subfloat[Firm type]{

\begin{tabular}{rll}
  \hline
 & firmName & firmID \\ 
  \hline
1 & MicroCapital Ltd & 123 \\ 
   \hline
\end{tabular}
}%
  \qquad
  \subfloat[Employee type]{

\begin{tabular}{rll}
  \hline
 & firstName & secondName \\ 
  \hline
1 & John & Smith \\ 
  2 & Peter & Pan \\ 
   \hline
\end{tabular}
  }
  
   \subfloat[Shareholder type]{

\begin{tabular}{rll}
  \hline
 & ID & Name \\ 
  \hline
1 & S1 & Karl Marx \\ 
  2 & S2 & Bill Gates \\ 
   \hline
\end{tabular}
  }
  
  \caption{The same data as in the XML code example above but mapped to data-frames according to the generic \pkg{RWebData} mapping procedure. While (a) contains data describing the firm (the main observational unit) itself, (b) and (c) contain the observations and variables describing the observation types employee and shareholder, respectively.}%
  \label{tab:mapping}%
\end{table}

\subsection{Parsing and generic mapping of different web data formats}
In order to map nested web data to data-frames, \pkg{RWebData} applies, in a first step, existing parsers for different formats (mime-types) of the data in the body of the HTTP response from the API to read the data into \proglang{R}.\footnote{In the case of JSON, either the parser provided in \citet{jsonlite_2014} or (in order to increase robustness in special cases) the one contributed by \citet{rjsonio_2013} is applied. XML-documents or RSS-documents are parsed with the parser provided in \citet{xml_2013} and YAML-documents with the parser provided in \citet{yaml_2014}.} Independent of the initial format, the data is parsed and coerced to a (nested) list representing the tree-structure of the raw data. In a next step, the data is mapped to one or several data-frames depending on the nesting structure and the recurrence of observation types and variables. Appendix~\ref{sec:rwebdata_A_algo} presents a detailed description of the mapping algorithm.  The data-frames are returned to the user either directly (in a list) or as part of an \code{apiresponse}-object. 

The core idea behind the generic approach to web data conversion in the \pkg{RWebData} package is thus to disentangle the parsing of web data from the mapping of the data to data-frames. This allows the parsers for different web data formats to be relatively simple, while the same mapping algorithm is focused on the data semantics and can be  applied independent of the initial raw data format. In addition, it allows different parsers' advantages to be combined in order to make the parsing process more robust. The suggested procedure goes hand in hand with the object-oriented approach applied in \pkg{RWebData} and provides summary and plot methods that work independent of the initial data format. The modular design of mapping web data to data-frames facilitates the extension of \pkg{RWebData}'s compatibility with additional web data formats or alternative parsers that are, for example, optimized for vast web documents. Parsers simply need to read the web data as (nested) lists or in a format that conserves the tree-structure of the raw web data and can be easily coerced to a nested list.

\subsection{Basic data mapping algorithm}

Once the web data-document is downloaded from an API and coerced to a nested list, the algorithm consists essentially of two procedures:
\begin{enumerate}
      \item The extraction of different observation types (and their respective observations and variables) as sub-trees. While reversely traversing the whole data-tree, the algorithm checks at each level (node) of the tree whether either the whole current sub-tree or one of its siblings can be considered an observation type. If so, the respective sub-tree is extracted and saved in a deposit variable. If not, the algorithm traverses further down in the tree-structure. Checks for observation types are defined as a set of control statements that incorporate the above outlined data semantics. Essentially, the recurrence of the same variables as siblings (containing the actual data values as leaf elements) is decisive in order to recognize observation types. The result of this step is a set of sub-trees, each sub-tree containing one observation type and its respective observations and variables.
      \item For each of the resulting observation types (in the form of sub-trees), the respective observations are extracted as character vectors and bound as rows in a data-frame, while the variable names are conserved and added as column names. This procedure is built to be robust to observations described by a differing number of variables. The result of this step is a set of data-frames, each describing one observation type and containing individual observations as rows and variables as columns.
\end{enumerate}

Singly occurring leaf nodes that are not nested within one of the detected observation types are collected along the way and then returned in one data-frame. According to the logic of the data semantics and the algorithm outlined above, these remaining data can be seen as metadata describing the data set as a whole. Appendix~\ref{sec:rwebdata_A_algo} presents a detailed, more technical outline of the data mapping algorithm.

\subsection{Basic architecture}

\begin{center}
\begin{figure}[h]
  \includegraphics[width=0.85\textwidth]{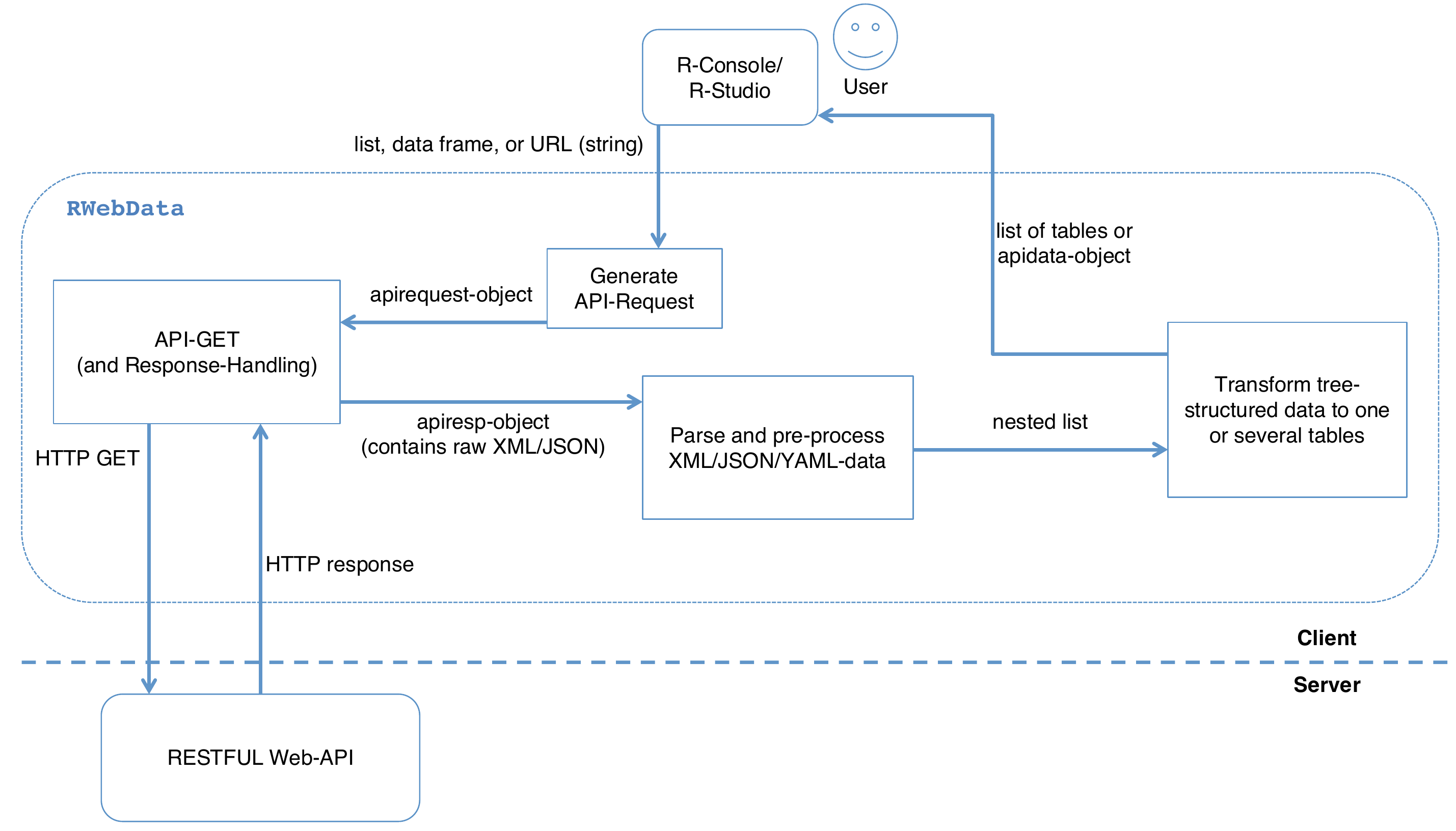}
\caption{Basic architecture of the \pkg{RWebData} package.}
\label{fig:architecture}
\end{figure}
\end{center}

\pkg{RWebData} is specifically written to give practical researchers a high-level interface to compile data from the programmable web. The common user will thus normally not be confronted with the range of internal processes outlined above. The work-flow with \pkg{RWebData} for an average user thus mainly involves the specification of what data should be requested from what API (in the form of either a list, a data-frame or simply a string with a URL) to one of \pkg{RWebData}'s high-level functions in order to obtain the desired data mapped to data-frames. Figure~\ref{fig:architecture} illustrates the basic architecture of \pkg{RWebData}, summarizing its internal functionality and the processing procedure when the package is used by an average user.

\pkg{RWebData}'s high level functions take either lists, data frames, or a URL (as character string) as input values and return a list of data-frames or an apidata-object. First, \pkg{RWebData} generates an \code{apirequest-object} based on which the HTTP GET and responses are handled. Upon successful interaction with the API, an \code{apiresponse-object}, containing, inter alia, the raw web data, is generated. Subject to consideration of the mime-type of the raw data, the data is then preprocessed, parsed, and coerced to a nested list. Finally, the data mapping algorithm is applied in order to extract the data in the form of data frames as outlined in the previous section. The latter builds the main part of the package. The procedure is straight-forward and user-friendly in practice.

\section[Basic functionality]{Basic functionality}
\label{sec:rwebdata_basics}

In order to describe the basic usage of \pkg{RWebData}, this section offers a step-by-step introduction to the high-level functionality of the package. Some of the functions described here offer more options which are not all discussed in this section.\footnote{Details on all these options are provided by the respective \proglang{R} help files. Users are generally encouraged to read the detailed package documentation of \pkg{RWebData}.}.

\pkg{RWebData}'s implementation is motivated by the convention over configuration paradigm and thus provides a way for the user to interact with APIs using as few specifications as necessary to access the data. \pkg{RWebData} then converts data in the most common formats provided by APIs to one data-frame or a list of several data-frames. Hence, the user does not need to specify any HTTP options or understand what XML or JSON is and is able to obtain the data directly in the expected format for statistical analysis. There are primarily two high-level functions in \pkg{RWebData} that provide this functionality in different ways: \code{getTabularData()} and \code{apiData()}. 

\subsection{Fetching data from REST APIs}

The function \code{getTabularData()} provides a straightforward way to get data from web APIs as data-frames. In the simplest case, the function takes a string containing the URL to the respective API resource as an input.\footnote{The URL to the resource which a user wants to query is very easy to find in any REST API documentation. How to call the respective resource methods of an API with URLs is the essential part of such documentations.} The function then handles the HTTP request and response, parses the body of the response (depending on the mime-type) and automatically extracts the data records from the nested data structure as data-frames. This enables us to fetch data from different APIs providing data in different formats with essentially the same command.

Consider, for example, the World Bank Indicators API which provides time series data on financial indicators of different countries.\footnote{Note that this is an example of an API which is -- unlike most APIs -- explicitly made for researchers retrieving data from the web. Nevertheless, retrieving data from this API with functions from, e.g., the \pkg{XML} package \citep{xml_2013} is not necessarily straightforward for users without a background in web technologies.} We want to use data from that API to investigate how the United States' public dept was affected by the financial crisis in 2008. All we need in order to download and extract the data is the URL to the respective resource on the API.\footnote{A URL to an API resource typically includes the base address of the API, a part specifying the specific API query method or resource, and some query parameters. How to build query-URLs in the specific case of the World Bank Indicators API is well documented on \url{http://data.worldbank.org/node/203}.}

\begin{Schunk}
\begin{Sinput}
R> u <- paste0("http://api.worldbank.org/countries/USA/indicators", # address
+            "/DP.DOD.DECN.CR.GG.CD?", # query method
+            "&date=2005Q1:2013Q4") # parameters
R> usdept <- getTabularData(u)
\end{Sinput}
\end{Schunk}

\begin{figure}[h]
 \centering
\includegraphics{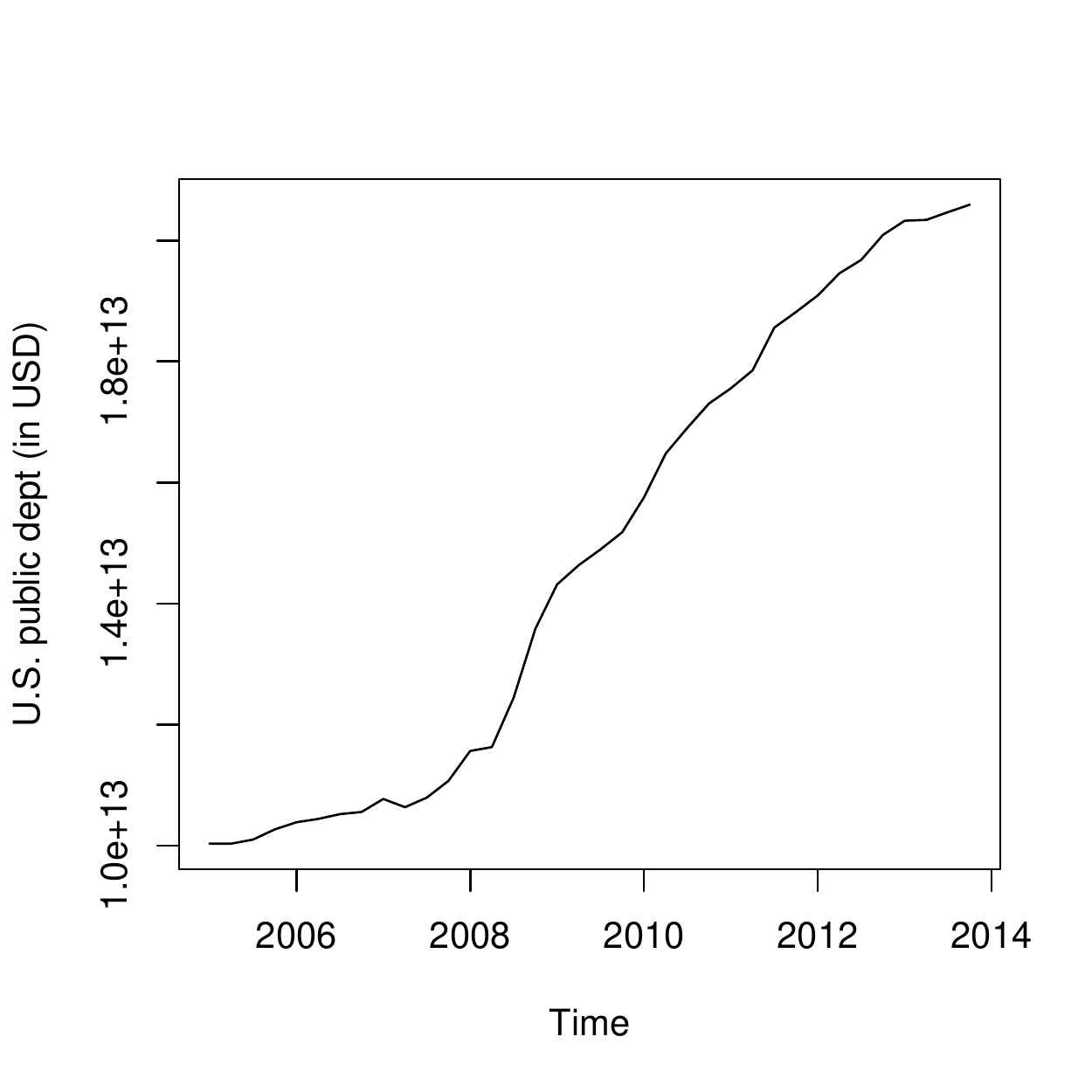}
\caption{Plot of the time series on United States' public dept extracted from the World Bank Indicators API}
\label{fig:wb}
\end{figure}

Without bothering about the initial format\footnote{The World Bank Indicators API provides time series data by default as XML in a compressed text file. Handling solely one API query of this type might already involve many steps to fetch and extract the data with the existing lower level functions. And, thus, might be tedious for a user who only has limited experience with web technologies.}, the returned data is already in the form of a data-frame and is ready to be analyzed (e.g., by plotting the time series as presented in Figure~\ref{fig:wb}):

\begin{Schunk}
\begin{Sinput}
R> require(zoo)
R> plot(as.ts(zoo(usdept$value,  as.yearqtr(usdept$date))), 
     ylab="U.S. public dept (in USD)")
\end{Sinput}
\end{Schunk}

The same approach of fetching data with \code{getTabularData()} can be applied to any other REST API that provides data in either XML/RSS, JSON, or YAML. Instead of a URL, one can also use a list of request parameters plus the base URL that directs to the API's server as function arguments. We use this approach in order to download data from the HarvardEvents API (\url{https://manual.cs50.net/api/events/}), which provides data about current or past events at Harvard University. Data on events can be fetched in various formats. In the following example, we query the API for upcoming lectures at Harvard and request a different data format each time (note the different format in the \code{output} parameter). 

\begin{Schunk}
\begin{Sinput}
R> hae_api <- "http://events.cs50.net/api/1.0/events?"
R> haexml <- getTabularData(list(output="xml", q="lecture"), base.url=hae_api)
R> haejson <- getTabularData(list(output="json", q="lecture"), base.url=hae_api)
R> haerss <- getTabularData(list(output="rss", q="lecture"), base.url=hae_api)
\end{Sinput}
\end{Schunk}

In each case, \code{getTabularData()} automatically converts the raw web data into data-frames.\footnote{Although an API offers its data in different formats, it does not necessarily send exactly the same content for the same request in different formats. Different formats might serve different purposes for web developers. It thus makes sense that some contain additional information. Therefore, data requests with \code{getTabularData()} to the same API but in different formats do not automatically lead to identical results. While \code{haexml} and \code{haejson} are almost identical in the example above, \code{haerss} differs from the other results.} The output to this example is presented in Appendix~\ref{sec:rwebdata_A_harvard}.  

While the data format in the examples above varies, the underlying data structure is relatively simple in all these examples. The World Bank Indicators API is specifically devised to provide data for statistical analysis. The HarvardEvents API, although primarily designed for developers, provides data that can be naturally represented in one table (with the exception of responses in RSS format). The next section considers examples where the provided data cannot easily be thought of as one single table.

\subsection{Nested data structures}

The high-level function \code{getTabularData()} automatically handles nested data and converts them to a list of various data-frames. It does not, however, provide any information on the initial nesting structure and can only handle individual requests. Alternatively, we can use \code{apiData()} in order to exploit \pkg{RWebData}'s internal classes and methods to handle requests to APIs. The simplest way to call \code{apiData()} is again by using a URL (as a string) pointing to an API resource. \code{apiData()} returns an object of class \code{apiresponse}. Such \code{apiresponse} objects contain additional information about the executed request to the API and support generic plot and summary functions illustrating the structure and content of the retrieved web document. The following example demonstrates the summary methods for \code{apiresponse} objects with data from the Ergast API (\url{http://ergast.com/mrd/}) on Formula 1 race results.  

\begin{Schunk}
\begin{Sinput}
R> f1 <- apiData('http://ergast.com/api/f1/2013/1/results.json',
+               shortnames=TRUE)
R> summary(f1)
\end{Sinput}
\begin{Soutput}
API data summary: 
=================

The API data has been split into the following 2 data frames:

         Length Class      Mode
metadata 20     data.frame list
Results  27     data.frame list

The respective data frame(s) contain the following variables:

1. metadata:
xmlns,  series,  url,  limit,  offset,  total,  season,  round,  1,  
raceName,  circuitId,  2,  circuitName,  lat,  long,  locality,  
country,  date,  time,  path, 

2. Results:
number,  position,  positionText,  points,  driverId,  
permanentNumber,  code,  url,  givenName,  familyName,  dateOfBirth,  
nationality,  constructorId,  url,  name,  nationality,  grid,  
laps,  status,  millis,  time,  rank,  lap,  time,  units,  speed,  
path, 
\end{Soutput}
\end{Schunk}

The summary method called by the generic \code{summary()}, provides an overview of the variables included in the data and shows how \pkg{RWebData} has split the data into several data-frames. This is particularly helpful in an early phase of a research project when exploring an API for the first time.

\begin{center}
\begin{figure}[h]
\includegraphics{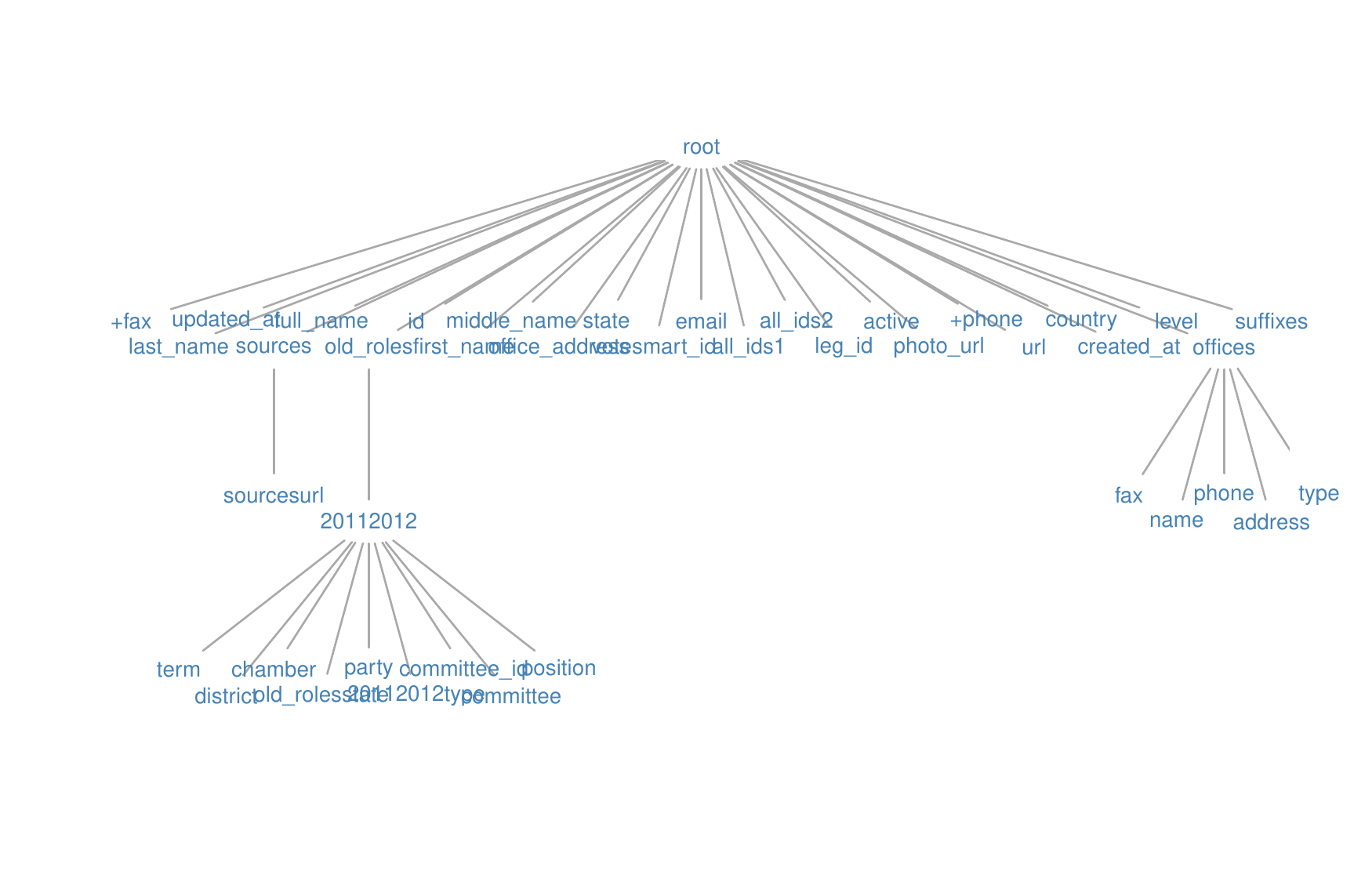}
\caption{Reingold-Tilford tree-graph illustrating the nested structure of the raw JSON data from the Open States API.}
\label{fig:pgraph}
\end{figure}
\end{center}

The next example demonstrates the visualization of nested data returned from the Open States API\footnote{See \url{https://sunlightlabs.github.io/openstates-api/} for details.}. We query data on a legislator in the Council of the District of Columbia with \code{apiData()} and call the generic \code{plot()} on the returned \code{apiresponse} object. Figure~\ref{fig:pgraph} shows the result of the plot command below. Note that the Open States API is free to use for registered users. Upon registration a api-key is issued that is then added as a parameter-value to each API request (indicated with ``\code{[YOUR-API-KEY]}" in the example below).

\begin{Schunk}
\begin{Sinput}
R> url <- "http://openstates.org/api/v1/legislators/DCL000004/
+  ?apikey=[YOUR-API-KEY]"
R> p <- apiData(url)
R> plot(p, type="jitter")
\end{Sinput}
\end{Schunk}

The \code{apiresponse} plot method illustrates the nesting structure of the original data by drawing entities (variables or types) as nodes and nesting relationships as edges in a Reingold-Tilford tree-graph (\citealt{reingold_tilford1981}; see also \citealt{igraph_2006} for the implementation in \proglang{R} on which \pkg{RWebData} relies.). The option \code{type="jitter"} shifts the nodes slightly vertically to make long node names better readable in the graph.

The transformed data (in the form of a list of data-frames) saved in an \code{apiresponse}-object can be accessed with \code{getdata()}. In the current example, the data has been split into two data-frames: one with metadata containing general variables describing the legislator, and one containing data on the legislator's roles in the 2011-2012 session.

\begin{Schunk}
\begin{Sinput}
R> pdata <- getdata(p)
R> summary(pdata)
\end{Sinput}
\begin{Soutput}
          Length Class      Mode
metadata  23     data.frame list
2011-2012 11     data.frame list
\end{Soutput}
\end{Schunk}

\subsection{Interactive sessions}

So far, we have only considered individual requests to APIs. In practice, users might want to query an API in many ways in an interactive session to explore the data and assemble the data set that meets their needs. This can easily be done with \pkg{RWebData}. We continue with the Open States API example. First, we fetch general data on all legislators in the DC Council.

\begin{Schunk}
\begin{Sinput}
R> url <- "http://openstates.org/api/v1/legislators/?state=dc&chamber=upper
+  &apikey=[YOUR-API-KEY]"
R> dc_council <- getTabularData(url)
\end{Sinput}
\end{Schunk}

The goal is to combine these general data with data on the legislators' roles. In particular, we want to download the detailed role data on all legislators and combine these within one data set. As there is no API method provided to obtain these data with one call to the API, we use \code{apiDownload()} to handle all individual requests at once. By inspecting the \code{dc_council} data-frame from above, we see that the fifth column of that data-frame contains all the IDs (id) pointing to the respective resources on individual council members. We simply paste these ids with the \code{legislators}-method and use the resulting URLs as the function argument to \code{apiDownload()} in order to obtain all the data compiled in one data-frame.

\begin{Schunk}
\begin{Sinput}
R> head(names(dc_council))
\end{Sinput}
\begin{Soutput}
[1] "fax"        "last_name"  "updated_at" "full_name"  "id"        
[6] "first_name"
\end{Soutput}
\begin{Sinput}
R> api_requests <- paste0("http://openstates.org/api/v1/legislators/", 
+                        dc_council$id,
+                        "/?apikey=",
+                        "YOUR-API-KEY")
\end{Sinput}
\end{Schunk}

\begin{Schunk}
\begin{Sinput}
R> dc_leg_roles <- apiDownload(api_requests)
\end{Sinput}
\end{Schunk}

During the download, \code{apiDownload()} indicates the number of queries processed on a progress bar printed to the R console. Moreover, \code{apiDownload()} periodically saves the processed data locally in a temporary file. This relieves the working memory assigned to the \proglang{R}-session and makes large downloads robust to network interruptions. A summary of the resulting \code{dc_leg_roles} is presented in Appendix~\ref{sec:rwebdata_A_openstates}.

\newpage 

\section{Writing interfaces to REST APIs}
\label{sec:rwebdata_advanced}

In the context of an ongoing research project, it might be more comfortable to have a specific function for specific queries to the same API.
We could manually program such a function based on \pkg{RWebData}'s internal functions or using the functions of other packages. However, \pkg{RWebData} provides a way to write such functions automatically. Given the parameters and the base URL for the respective API method/resource, \code{generateQueryFunction()} writes a function that includes all the functionality of the package to handle queries to that specific API method/resource. The same works for APIs that accept requests via form URLs, such as the BisTip.com Search API. Here, only the base URL and the respective parameters have to be specified.

\begin{Schunk}
\begin{Sinput}
R> bi_trip <- "http://www.bistip.com/api/v1/trips.json?"
R> bi_seek <- "http://www.bistip.com/api/v1/seeks.json?"
R> bi_params <- list(from=NA, to=NA) # parameters without default value
R> bitstipTrip <- generateQueryFunction(x=bi_params, base.url=bi_trip)
R> bitstipSeek <- generateQueryFunction(x=bi_params, base.url=bi_seek)
\end{Sinput}
\end{Schunk}

The new functions \code{bitstipTrip()} and \code{bitstipSeek()} can now be used to query the BitsTip.com API. 

\begin{Schunk}
\begin{Sinput}
R> all_t <- bitstipSeek(from="tokyo")
R> j_s <- bitstipTrip(from="jakarta", to="singapore")
R> names(j_s)[1:4]
\end{Sinput}
\begin{Soutput}
[1] "departure_date_medium_format" "origin_location"             
[3] "arrival_date_medium_format"   "period"                      
\end{Soutput}
\end{Schunk}

With only five lines of code, we have generated a complete BitsTip.com API \proglang{R}-client library (or Open Source Interface; \citealt{matter_stutzer2015plos}) that automatically provides the web data as data-frames. 

\section[Discussion]{Discussion}
\label{sec:rwebdata_discussion}

Empirically driven social sciences can substantially profit from the rapidly growing programmable web as a data source covering countless dimensions of socio-economic activity. However, the purpose and goals that drove the initial development of web APIs focused on providing general data for public use via dynamic websites and other applications, and not on the provision of data formatted for scientific research and statistical analyses. This leads to technical hurdles a researcher has to overcome in order to compile such data in a format that is suitable for statistical analysis. The presented \proglang{R} package \pkg{RWebData} suggests a simple high-level interface that helps to overcome such technical hurdles (and the respective costs) associated with the statistical analysis of data from the programmable web. The package contributes to a frictionless and well documented raw data compilation and data preparation process that substantially increases the replicability and reproducibility of original research based on data from the programmable web. As pointed out by \cite{crosas_etal2015}, social science research with big data poses generally new challenges to the reuse of data as well as the reproduction of results. Empirical research based on newly available big public data from web data sources is a challenge demanding scientific rigor. The need for replicability is even greater in the age of big data: Concision must be an essential characteristic of data recording and retrieval with regard to parsing and compilation. Rigor applies not only to the final prepared data used in the analysis, but also to the documentation of the data preparation and data selection process. With the further development of \pkg{RWebData}, an R-script documenting how the package's high-level functions have been applied to compile data as well as any further step in the data preparation and analysis are enough to ensure the replicability of a study based on big public data from the programmable web.

\section*{Acknowledgements}
I am grateful to Dietmar Maringer, Armando Meier, Reto Odermatt, Michaela Slotwinski, Alois Stutzer, as well as seminar participants at the University of Basel and the University of Oxford for helpful remarks. Special thanks go to Ingmar Schlecht for many productive discussions on software development and the methodological aspects of this paper. I also thank Joerg Kalbfuss for excellent research assistance. The author acknowledges financial support from the University of Basel Research Fund.

\newpage

\bibliography{rwebapi}

\newpage
\setcounter{table}{0}
\renewcommand{\thetable}{A\arabic{table}}
\renewcommand{\thealgorithm}{A\arabic{algorithm}}
\renewcommand{\thesubsection}{A.\Roman{subsection}}

\section*{Appendix}

\subsection{Data mapping algorithm}
\label{sec:rwebdata_A_algo}

As pointed out in the main text, \pkg{RWebData} parses and coerces the raw web data to a nested list representing the tree-structure of the data. Call this list $\vec{x}$.

The data mapping algorithm consists of two main parts. First, $\vec{x}$ is split into $n$ lists representing sub-trees. One for each \emph{observation type} in the data. The key problem that the algorithm has to solve at this step is the identification of cutting points (i.e., what part of the tree belongs to what observation type). Second, each resulting sub-tree is then split into individual character vectors. One for each \emph{observation}. The individual observations are then stacked together in one data-frame with each vector (observation) as a row. Algorithm~\ref{algo:types_observations} presents a formal description of these procedures. In the resulting data-frames, each row $i$ represents one \emph{observation}, and each column $j$ represents a \emph{variable/characteristic} of the $n$ observations. The data-frames are then returned in a list. 

In order to make the formal descriptions of the procedures in the data mapping algorithm conveniently readable, the pseudo-code describes simple versions of these procedures (that are not necessarily most efficient). The algorithms' \proglang{R}-implementations in \pkg{RWebData} are more efficient and contain more control statements to ensure robustness. The actual \proglang{R}-implementation relies partly on existing \proglang{R} functions and favors  vectorization over for-loops in some cases.

\begin{algorithm}
\caption{Data mapping algorithm}
\label{algo:types_observations}
\begin{algorithmic}[1]

\Procedure{Types}{$\vec{x}$}
\State $Types \gets$ empty list
\State $deposit \gets$ empty list
\If{$\vec{x}$ is an observation type}
      \State add $\vec{x}$ to $Types$
      \State \textbf{return} $Types$
\EndIf
\If{$\vec{x}$ contains a part $i$ at the highest nesting level that is an observation type}
      \State add $i$ to $Types$
      \State remove $i$ from $\vec{x}$
\EndIf
\ForAll{elements $i$ in $\vec{x}$}
\If{$i$ is a non-empty list}
      \If{$i$ is an observation type}
            \State add $i$ to $Types$
      \Else
            \State apply this very procedure to $i$ \Comment{recursive call}
            \State add the resulting observation types to $Types$
            \State add the remaining leaves (metadata) to $deposit$
      \EndIf
\Else
      \State add $i$ to $deposit$ \Comment{it's a leaf node (i.e., just a value)}
\EndIf
\EndFor
\EndProcedure

\Procedure{Observations}{$Types$}
\State $rows \gets$ empty list

\ForAll{$\vec{obstype}$ in $Types$}
    \ForAll{$observation$ in $\vec{obstype}$}
    \State unlist and transpose $observation$ \Comment{extract leaves as vector}
    \State add resulting vector to $rows$ 
    \EndFor
    \State bind rows to one data-frame (preserving variable names)
\EndFor 
  
\EndProcedure
\end{algorithmic}
\end{algorithm}

\newpage

\subsection{HarvardEvents API example }
\label{sec:rwebdata_A_harvard}

\begin{Schunk}
\begin{Sinput}
R> hae_api <- "http://events.cs50.net/api/1.0/events?"
R> haexml <- getTabularData(list(output="xml", q="lecture"), base.url=hae_api)
R> haejson <- getTabularData(list(output="json", q="lecture"), base.url=hae_api)
R> haerss <- getTabularData(list(output="rss", q="lecture"), base.url=hae_api)
R> # comparison of results based on xml/json data:
R> dim(haexml)
\end{Sinput}
\begin{Soutput}
[1] 14 10
\end{Soutput}
\begin{Sinput}
R> names(haexml)
\end{Sinput}
\begin{Soutput}
 [1] "summary"        "dtstart"        "dtend"         
 [4] "location"       "description"    "calname"       
 [7] "id"             "calname"        "INPUT_:_output"
[10] "INPUT_:_q"     
\end{Soutput}
\begin{Sinput}
R> haexml[1,4]
\end{Sinput}
\begin{Soutput}
[1] "The Harvard Ed Portal, 224 Western Ave., Allston, Mass."
\end{Soutput}
\begin{Sinput}
R> dim(haejson)
\end{Sinput}
\begin{Soutput}
[1] 14 10
\end{Soutput}
\begin{Sinput}
R> names(haejson)
\end{Sinput}
\begin{Soutput}
 [1] "summary"        "dtstart"        "dtend"         
 [4] "location"       "description"    "calname"       
 [7] "id"             "calname"        "INPUT_:_output"
[10] "INPUT_:_q"     
\end{Soutput}
\begin{Sinput}
R> haejson[1,4]
\end{Sinput}
\begin{Soutput}
[1] "The Harvard Ed Portal, 224 Western Ave., Allston, Mass."
\end{Soutput}
\begin{Sinput}
R> # rss response is differently structured:
R> summary(haerss)
\end{Sinput}
\begin{Soutput}
         Length Class      Mode
metadata 8      data.frame list
item     8      data.frame list
\end{Soutput}
\begin{Sinput}
R> haerss$item[1,]
\end{Sinput}
\begin{Soutput}
                             guid                            title
1 http://events.cs50.net/20281645 Fundamentals of Social Media 1.0
                             link description category
1 http://events.cs50.net/20281645        <NA>   events
                          pubDate INPUT_:_output INPUT_:_q
1 Tue, 19 Jul 2016 17:00:00 -0400            rss   lecture
\end{Soutput}
\begin{Sinput}
R> 
R> 
\end{Sinput}
\end{Schunk}

\subsection{Open States API}
\label{sec:rwebdata_A_openstates}

\begin{Schunk}
\begin{Sinput}
R> url <- "http://openstates.org/api/v1/legislators/?state=dc&chamber=upper
+  &apikey=[YOUR-API-KEY]"
R> c_council <- getTabularData(url)
R> api_requests <- paste0("http://openstates.org/api/v1/legislators/", 
                       dc_council$id,
                       "&apikey=",
                       "YOUR-API-KEY")
\end{Sinput}
\end{Schunk}

\begin{Schunk}
\begin{Soutput}
  |                                                                  
  |                                                            |   0%
  |                                                                  
  |==============================                              |  50%
  |                                                                  
  |============================================================| 100%
\end{Soutput}
\end{Schunk}

\begin{Schunk}
\begin{Sinput}
R> # explore the data
R> summary(dc_leg_roles)
\end{Sinput}
\begin{Soutput}
          Length Class      Mode
metadata  54     data.frame list
2013-2014 13     data.frame list
2011-2012 13     data.frame list
\end{Soutput}
\begin{Sinput}
R> dim(dc_leg_roles$roles) 
\end{Sinput}
\begin{Soutput}
NULL
\end{Soutput}
\end{Schunk}

\end{document}